\documentclass[pra,showpacs]{revtex4}
\usepackage{amsfonts}

\begin{document}

\title{Applications of Canonical Transformations}
\author{Subhashish Banerjee and Joachim Kupsch\cite{rg}}
\affiliation{Fachbereich Physik, TU Kaiserslautern, D-67653 Kaiserslautern,
Germany}
\begin{abstract}
Canonical transformations are defined and discussed along with the
exponential, the coherent and the ultracoherent vectors. It is shown that
the single-mode and the $n$-mode squeezing operators are elements of the
group of canonical transformations. An application of canonical
transformations is made, in the context of open quantum systems, by studying
the effect of squeezing of the bath on the decoherence properties of the
system. Two cases are analyzed. In the first case the bath consists of a
massless bosonic field with the bath reference states being the squeezed
vacuum states and squeezed thermal states while in the second case a system
consisting of a harmonic oscillator interacting with a bath of harmonic
oscillators is analyzed with the bath being initially in a squeezed thermal
state.
\end{abstract}

\pacs{03.65.-w, 03.65.Yz, 05.30.-d}

\maketitle

\section*{I. Introduction}

The transformations preserving the form of the Hamilton equations are known
as the canonical transformations in classical mechanics. Their counterparts
in quantum mechanics are those that preserve the commutation relations
between the creation and the annihilation operators. Many physical
consequences can be derived from canonical transformations in quantum
mechanics \cite{kn90}. For instance, in the phase-space picture the
uncertainty relations are formulated in terms of the area occupied by the
Wigner function \cite{ho84}. The spread of the wave packet is an
area-preserving canonical transformation in quantum mechanics. The Lorentz
boost in a given direction is a canonical transformation in phase space
using the light-cone variables. This allows the formulation of the
uncertainty relations in a Lorentz-invariant manner.

Decoherence is a consequence of the `openness' of a physical system \cite
{fv63, cl83, jz03, bg03}, i.e., the system is not isolated but is in contact
with its environment. This causes the decay of the quantum interferences
between component states of quantum superposition states which quantify the
nonclassical effects in quantum mechanics. From this and also from the point
of view of quantum computation it would be very desirable to have some means
of stabilizing the quantum superpositions against action of the environment
\cite{vl98, ss97, zr97, vt01, ga99}. In this context, a number of studies
have been made \cite{kw88, kb93} which have shown that the phase-sensitive
squeezed reservoirs (environment) are better suited to the task of
controlling decoherence than the phase-insensitive thermal heat bath
(reservoir). Squeezed baths are characterized by the fact that correlations
exist between the bath modes \cite{cc82}.

In this paper we will discuss some applications of the generalized canonical
transformations the details of which have been formulated in \cite{kb04}.
The plan of the paper is as follows. In Section II we will review canonical
transformations illustrating its group structure and discuss the
exponential, the coherent and the ultracoherent vectors as well as the
canonical transformations in Fock space. In Section III we will demonstrate
the single-mode and the $n$-mode squeezing operators to be elements of the
group of canonical transformations and thus apply their unitary ray
representations to the coherent as well as the ultracoherent states. In
Section IV as an application of squeezing to open quantum systems, we
discuss the effect of squeezing of the bath on the decoherence properties of
the system. In Section IV(A) we discuss the induced superselection rules of
a class of Hamiltonian models with the environment given by a massless
bosonic field and the environment reference states being squeezed vacuum and
squeezed thermal states. In Section IV(B) we make use of the squeezed
thermal state on a generic model of open quantum systems and demonstrate
its ability to put a check on the decoherence properties of the system. In
Section V we make our conclusions. We have also included an appendix in
which we discuss some details of the ultracoherent vector and its connection
with canonical transformations.

\section*{II. Ultracoherent vectors and canonical transformations}

In this section we recapitulate the properties of canonical transformations
on a Fock space using coherent and ultracoherent vectors. A more detailed
presentation of this technique will be given in \cite{kb04}, see also
Appendix A of the present paper. We start with some basic definitions and
notations about the Fock space of symmetric tensors. Let $\mathcal{H}$ be
Hilbert space with inner product $\left( f\mid g\right) $ and with an
antiunitary involution $f\rightarrow f^{\ast }$. The mapping $%
f,\,g\rightarrow \left\langle f\mid g\right\rangle :=\left( f^{\ast }\mid
g\right) \in \mathbb{C}$ is then a symmetric bilinear form. An explicit and
representative example is $\mathcal{H=}\mathbb{C}^{n}$ with $\left( f\mid
g\right) =\sum_{\mu =1}^{n}\overline{f}_{\mu }g_{\mu }$ and $\,\left\langle
f\mid g\right\rangle =\sum_{\mu =1}^{n}f_{\mu }g_{\mu }$ for $f=\left(
f_{1},...,f_{n}\right) \in \mathbb{C}^{n}$ and $g=\left( g_{1},...,g_{n}\right)
\in \mathbb{C}^{n}$. The involution is simply complex conjugation $f^{\ast }=$ $%
\left( \overline{f}_{1},...,\overline{f}_{n}\right) $. With the symmetric
tensor product $f\circ g$ the Hilbert space $\mathcal{H}$ generates the
bosonic Fock space $\mathcal{F}(\mathcal{H})$. The vacuum vector is denoted
by $1_{vac}$. For all vectors $f\in \mathcal{H}$ the exponential vectors $%
\exp f=1_{vac}+f+\frac{1}{2}f\circ f+...$ converge within $\mathcal{F}(%
\mathcal{H})$, the inner product being $\left( \exp f\mid \exp g\right)
=\exp \left( f\mid g\right) $. Coherent states are the normalized
exponential vectors $\exp \left( f-\frac{1}{2}\left\| f\right\| ^{2}\right) $%
. The linear span of all exponential vectors $\left\{ \exp f\mid f\in
\mathcal{H}\right\} $ is dense in $\mathcal{F}(\mathcal{H})$. To determine
an operator on $\mathcal{F}(\mathcal{H})$ it is therefore sufficient to know
this operator on all exponential vectors. The involution on $\mathcal{H}$ is
naturally extended to an involution on $\mathcal{F}(\mathcal{H})$ such that $%
\left( \exp f\right) ^{\ast }=\exp f^{\ast },\,f\in \mathcal{H}$. The
mapping $f,\,g\in \mathcal{F}(\mathcal{H})\rightarrow \left\langle f\mid
g\right\rangle =\left( f^{\ast }\mid g\right) \in \mathbb{C}$ is again a
bilinear symmetric form.

A class of elements of the Fock space more general than coherent vectors are
the ultracoherent vectors \cite{sl88}. Let $A$ be a Hilbert-Schmidt operator
on $\mathcal{H}$, which is symmetric, i.e., $\left\langle Af\mid
g\right\rangle =\left\langle f\mid Ag\right\rangle $ holds for all $f,\,g\in
\mathcal{H}$, then there exists a unique tensor of second degree, in the
sequel denoted by $\Omega (A)$, such that
\begin{equation}
\left\langle \Omega (A)\mid f\circ g\right\rangle =\left\langle Af\mid
g\right\rangle  \label{2.1}
\end{equation}
for all $f,g\in \mathcal{H}$. To be more explicit, let $e_{\mu }=e_{\mu
}^{\ast }$ be a real orthonormal basis of the Hilbert space $\mathcal{H}$,
then $\Omega (A)$ is the tensor $\Omega (A)=\frac{1}{2}\sum_{\mu \nu }A_{\mu
\nu }\,e_{\mu }\circ e_{\nu }$ where the coefficients $A_{\mu \nu
}=\left\langle e_{\mu }\mid A\,e_{\nu }\right\rangle =A_{\nu \mu }$ have a
convergent sum $\sum_{\mu \nu }\left| A_{\mu \nu }\right| ^{2}<\infty $. If $%
A^{\dagger }A<I$ (all eigenvalues strictly less than one) then the norm of $%
\Omega (A)$ is strictly less than $1/\sqrt{2}$ and the exponential series $%
\exp \Omega (A)$ converges within the Fock space, the norm being $\left\|
\exp \Omega (A)\right\| =\det (I-A^{\dagger }A)^{-\frac{1}{4}}$. We denote
with $\mathcal{D}_{1}$ the set of all symmetric Hilbert-Schmidt operators
with $A^{\dagger }A<I$. The convex set $\mathcal{D}_{1}$ is usually called
the Siegel disc. For any operator $Z\in \mathcal{D}_{1}$ and for any $f\in
\mathcal{H}$ we now define the ultracoherent vector
\begin{equation}
\mathcal{E}\left( Z,f\right) =\exp \Omega (Z)\circ \exp f=\exp \left( \Omega
(Z)+f\right) \in \mathcal{F}(\mathcal{H}).  \label{2.2}
\end{equation}

The standard building blocks of a bosonic theory are the creation and
annihilation operators. Given a vector $f\in \mathcal{H}$ the creation
operator $b^{\dagger}(f)$ of that vector and the annihilation operator $b(f)$
are uniquely determined by
\begin{eqnarray}
b^{\dagger }(f)\exp g &=&f\circ \exp g=\frac{\partial }{\partial \lambda }%
\exp (g+\lambda f)\mid _{\lambda =0},  \label{2.3} \\
b(f)\exp g &=&\left\langle f\mid g\right\rangle \exp g= \left( f^{\ast }\mid
g\right)\exp g.  \label{2.4}
\end{eqnarray}
These operators satisfy $\left( b^{\dagger }(f)\right) ^{\dagger }=b(f^{\ast
})$, and they have the commutation relations $\left[ b^{\dagger
}(f),b^{\dagger }(g)\right] =\left[ b(f),b(g)\right] =0$ and $\left[
b(f),b^{\dagger }(g)\right] =\left\langle f\mid g\right\rangle $. If we
choose an orthonormal basis $e_{\mu },\,\mu =1,2,...$, of $\mathcal{H}$,
then the operators $b_{\mu }^{\dagger }=b^{\dagger }(e_{\mu })$ and $b_{\nu
}=b(e_{\nu }^{\ast })$ form a basis of the operator algebra and satisfy the
canonical commutation relations
\begin{eqnarray}
\left[ b_{\mu }^{\dagger },b_{\nu }^{\dagger }\right] &=&\left[ b_{\mu
},b_{\nu }\right] =0,  \label{2.5} \\
\left[ b_{\mu },b_{\nu }^{\dagger }\right] &=&\delta _{\mu \nu }.
\label{2.6}
\end{eqnarray}
Below we shall use a notation of \cite{Berezin:1966}. Let $B$ be a
self-adjoint operator on $\mathcal{H}$ and $e_{\mu }$ be a orthonormal basis
of the Hilbert space $\mathcal{H}$. Then
\begin{equation}
b^{\dagger }Bb=\sum_{\mu \nu }\left( e_{\mu }\mid Be_{\nu }\right) \,b_{\mu
}^{\dagger }b_{\nu }  \label{2a}
\end{equation}
is a well defined self-adjoint operator on the Fock space. The operators $%
b_{\mu }^{\dagger },\,b_{\nu }$ and the matrix elements $\left( e_{\mu }\mid
Be_{\nu }\right) $ depend explicitly on the choice of the basis, but (\ref
{2a}) does not. Let $A$ be a symmetric Hilbert-Schmidt operator then $%
b^{\dagger }Ab^{\dagger }$ and $b\overline{A}b$ are defined as
\begin{equation}
b^{\dagger }Ab^{\dagger }=\sum_{\mu \nu }\left\langle e_{\mu }\mid Ae_{\nu
}\right\rangle b_{\mu }^{\dagger }\,b_{\nu }^{\dagger }\;\mathrm{and}\;b%
\overline{A}b=\left( b^{\dagger }Ab^{\dagger }\right) ^{\dagger }=\sum_{\mu
\nu }\overline{\left\langle e_{\mu }\mid Ae_{\nu }\right\rangle }b_{\mu
}\,b_{\nu }.  \label{2b}
\end{equation}

For arbitrary elements $h\in \mathcal{H}$ the Weyl operators are defined on
the set of exponential vectors by
\begin{equation}
W(h)\exp f=\exp \left( -\left( h\mid f\right) -\frac{1}{2}\left\| h\right\|
^{2}\right) \exp (f+h).  \label{2.7}
\end{equation}
These operators are unitary with $W^{\dagger }(h)=W(-h)=W^{-1}(h)$. The
definition (\ref{2.3}) is equivalent to
\begin{equation}
W(h)=\exp \left( b^{\dagger }(h)-b(h^{\ast })\right) .  \label{2.8}
\end{equation}
The identities (\ref{2.3}) - (\ref{2.7}) imply
\begin{eqnarray}
W(h)b_{\mu }^{\dagger }W(h) &=&b_{\mu }^{\dagger }-\overline{h}_{\mu },
\label{2.9} \\
W(h)b_{\nu }W(h) &=&b_{\nu }-h_{\nu }.  \label{2.10}
\end{eqnarray}
Hence the Weyl operators are the displacement operators of quantum optics.
As already stated, the linear span of exponential vectors or coherent
vectors is dense in the Fock space. From (\ref{2.7}) immediately follows
that the Weyl operators map this set into itself. Moreover, the identity $%
W(h)1_{vac}=\mathrm{\exp }\left( h-\frac{1}{2}\left\| h\right\| ^{2}\right) $
implies that the linear span of $\left\{ W(h)1_{vac}\mid h\in \mathcal{H}%
\right\} $ is exactly the linear span of all coherent states. The action of
the Weyl operator on an ultracoherent vector is \cite{kb04}
\begin{equation}
W(h)\exp \left( \Omega (A)+f\right) =\mathrm{e}^{-\frac{1}{2}\left\|
h\right\| ^{2}+\frac{1}{2}\left\langle h^{\ast }\mid Ah^{\ast
}-2f\right\rangle }\exp \left( \Omega (A)+f+h-Ah^{\ast }\right) .
\label{2.10a}
\end{equation}
Hence the Weyl operators map ultracoherent vectors onto ultracoherent
vectors.

Canonical transformations are affine linear transformations between the
creation and annihilation operators preserving the commutation relations.
From (\ref{2.9}) and (\ref{2.10}) immediately follows that Weyl operators
generate inhomogeneous canonical transformations. Treating the most general
linear homogeneous transformations we have
\begin{equation}
b^{\dagger }(f)\rightarrow b^{\dagger }(Uf)-b(\overline{V}%
f),\,b(f)\rightarrow b(\overline{U}f)-b^{\dagger }(Vf).  \label{2.11}
\end{equation}
Here $U$ and $V$ are bounded operators on the Hilbert space $\mathcal{H}$.
The operators $\overline{U}$ (and $\overline{V}$) are defined by $\overline{U%
}f=\left( Uf^{\ast }\right) ^{\ast }$ and $U^{T}$ means $U^{T}=\left(
\overline{U}\right) ^{\dagger }=\overline{U^{\dagger }}$. In the case of $%
\mathcal{H}=\mathbb{C}^{n}$ the corresponding matrices are just the complex
conjugate or the transposed matrix. The transformation (\ref{2.11}) is
equivalent to the following transformation of the argument of the Weyl
operator (\ref{2.8})
\begin{equation}
b^{\dagger }(f)-b(f^{\ast })\rightarrow b^{\dagger }(Uf+Vf^{\ast })-b(%
\overline{U}f^{\ast }+\overline{V}f),  \label{2.12}
\end{equation}
which can be better visualized by the mapping of the test functions
\begin{equation}
\left(
\begin{array}{c}
f \\
f^{\ast }
\end{array}
\right) \rightarrow \left(
\begin{array}{c}
Uf+Vf^{\ast } \\
\overline{V}f+\overline{U}f^{\ast }
\end{array}
\right) =\widehat{G}\left(
\begin{array}{c}
f \\
f^{\ast }
\end{array}
\right)  \label{2.13}
\end{equation}
where the matrix of operators
\begin{equation}
\widehat{G}=\left(
\begin{array}{cc}
U & V \\
\overline{V} & \overline{U}
\end{array}
\right)  \label{2.14}
\end{equation}
maps the underlying real space of $\mathcal{H}$, parametrized by the vectors
$\left(
\begin{array}{c}
f \\
f^{\ast }
\end{array}
\right) ,\,f\in \mathcal{H}$, into itself. The transformations (\ref{2.11})
preserve the canonical commutation relations, if
\begin{equation}
\widehat{G}\Theta \widehat{G}^{\dagger }=\Theta  \label{2.15}
\end{equation}
and
\begin{equation}
\Delta \widehat{G}\Delta =\overline{\widehat{G}}  \label{2.16}
\end{equation}
with
\begin{equation}
\Theta =\left(
\begin{array}{cc}
I & 0 \\
0 & -I
\end{array}
\right) ,\quad \Delta =\left(
\begin{array}{cc}
0 & I \\
I & 0
\end{array}
\right) .  \label{2.17}
\end{equation}
From these, we can derive the equivalent conditions for the transformations
to be canonical as
\begin{equation}
UU^{\dagger }-VV^{\dagger }=I,\;UV^{T}=VU^{T},  \label{2.18}
\end{equation}
or
\begin{equation}
U^{\dagger }U-V^{T}\overline{V}=I,\;U^{T}\overline{V}=V^{\dagger }U.
\label{2.19}
\end{equation}

The operators (\ref{2.14}) form the group $\mathcal{G}_{c}$ of linear
canonical transformations, which are often called Bogoliubov transformations
\cite{Bog:1947}. Thereby it is sufficient to identify the mapping $G=G(U,V)$
in the first line of (\ref{2.13})
\begin{equation}
f\in \mathcal{H}\rightarrow G(U,V)f=Uf+Vf^{\ast }\in \mathcal{H},
\label{2.20}
\end{equation}
which is an $\mathbb{R}$-linear transformation on $\mathcal{H}$. The successive
application of canonical transformations corresponds to the multiplication
of the respective matrix operators (\ref{2.14}) or of the respective
operators (\ref{2.20}). In the latter case the multiplication law follows
from the definition as
\begin{equation}
G_{2}G_{1}f=G_{2}\left( U_{1}f+V_{1}f^{\ast }\right) =\left( U_{2}U_{1}+V_{2}%
\overline{V}_{1}\right) f+\left( U_{2}V_{1}+V_{2}\overline{U}_{1}\right)
f^{\ast }.  \label{2.21}
\end{equation}
The inverse mapping of (\ref{2.20}) is
\begin{equation}
G^{-1}(U,V)f=U^{\dagger }f-V^{T}f^{\ast }=G(U^{\dagger },-V^{T})f.
\label{2.20a}
\end{equation}
In the finite dimensional case $\mathcal{H}=\mathbb{C}^{n}$ the identity (\ref
{2.15}) implies that $\mathcal{G}_{c}$ is a subgroup of $SU(n,n)$. The
identity (\ref{2.16}) is an additional reality constraint, such that $%
\mathcal{G}_{c}$ is isomorphic to the real symplectic group \cite
{Bargmann:1970}.

For finite dimensional Hilbert spaces $\mathcal{H}$ the canonical
transformations (\ref{2.11}) can always be implemented by unitary operators
on the Fock space $\mathcal{F}(\mathcal{H})$; in the infinite dimensional
case one needs the additional constraint that $V$ is a Hilbert-Schmidt
operator \cite{Berezin:1966, KMTP:1967}.

In order to define canonical transformations in Fock space, we set up a
projective representation of the group $\mathcal{G}_{c}$ by identifying for
each element $G$ of $\mathcal{G}_{c}$ a unitary operator $T(G)$ on $\mathcal{%
F}(\mathcal{H}) $ such that
\begin{equation}
T(id)=I,\,T(G_{2})T(G_{1})=\omega (G_{2},G_{1})T(G_{2}G_{1})  \label{2.22}
\end{equation}
with a multiplier $\omega (G_{2},G_{1})\in \mathbb{C},\,\left| \omega
(G_{2},G_{1})\right| =1$. It is sufficient to define $T(G)$ on the set of
exponential vectors
\begin{equation}
T(G)\exp f=\det \left| U\right| ^{-\frac{1}{2}}\exp \left( \Omega
(U^{\dagger -1}V^{T})+U^{\dagger -1}f-\frac{1}{2}\left\langle f\mid
V^{\dagger }U^{\dagger }{}^{-1}f\right\rangle \right) .  \label{2.25}
\end{equation}
Thereby the operator $\left| U\right| =\sqrt{UU^{\dagger }}=\sqrt{%
I+VV^{\dagger }}\geq I$ is the positive self-adjoint part of $U$. Since $V$
is a Hilbert-Schmidt operator, we know that $\left| U\right| -I$ is a trace
class operator, and the determinant $\det \left| U\right| $ is well defined
also if $\dim \mathcal{H}=\infty $. The formula (\ref{2.25}) shows that in
general canonical transformations map coherent vectors -- including the
vacuum -- onto ultracoherent vectors. This class of vectors turns out to be
stable against canonical transformations, see Appendix A and \cite{kb04}. In
order to work out the group structure of $T(G)$, its action on the
ultracoherent vector (\ref{2.2}) is also needed. This is illustrated in
Appendix A and \cite{kb04}.

All canonical transformations are products of the following two classes of
canonical transformations, cf. \cite{Bargmann:1970}.

\begin{enumerate}
\item  Take a self-adjoint operator $\Psi =\Psi ^{\dagger }$ on $\mathcal{H}$%
. Then
\begin{equation}
\widehat{G}=\exp i\left(
\begin{array}{cc}
\Psi & 0 \\
0 & -\overline{\Psi }
\end{array}
\right) =\left(
\begin{array}{cc}
U & 0 \\
0 & \overline{U}
\end{array}
\right)  \label{2.26}
\end{equation}
is a matrix operator of the type (\ref{2.14}), where the unitary operator
\begin{equation}
U=\exp i\Psi  \label{2.27}
\end{equation}
and $V=0$ obviously satisfy the conditions (\ref{2.18}). The transformation (%
\ref{2.20}) $G(U,0)$ coincides with the unitary operator $U$. We simply
denote $T(G(U,0))$ by $R(U)$. From (\ref{2.25}) we obtain
\begin{equation}
R(U)\exp f=\exp Uf.  \label{2.28}
\end{equation}
In this case the homogeneous canonical transformations map coherent states
onto coherent states, and ultracoherent vectors are mapped onto
\begin{equation}
R(U)\exp \left( \Omega (Z)+f\right) =\exp \left( \Omega (UZU^{T})+Uf\right) .
\label{2.29}
\end{equation}

\item  As the second case of a canonical transformation take a symmetric
Hilbert-Schmidt operator $\Xi =\Xi ^{T}$ on $\mathcal{H}$. Then
\begin{equation}
\widehat{G}=\exp \left(
\begin{array}{cc}
0 & \Xi \\
\overline{\Xi } & 0
\end{array}
\right) =\left(
\begin{array}{cc}
U & V \\
\overline{V} & \overline{U}
\end{array}
\right)  \label{2.30}
\end{equation}
is a matrix operator of the type (\ref{2.14}) with the bounded operators
\begin{equation}
U=\cosh \sqrt{\Xi \overline{\Xi }}\geq I\quad \mathrm{and}\quad V=\Xi \frac{%
\sinh \sqrt{\overline{\Xi }\Xi }}{\sqrt{\overline{\Xi }\Xi }}=\frac{\sinh
\sqrt{\Xi \overline{\Xi }}}{\sqrt{\Xi \overline{\Xi }}}\Xi =V^{T},
\label{2.31}
\end{equation}
which satisfy the conditions (\ref{2.18}). Moreover, $U-I$ is a positive
trace class operator and $V$ is a Hilbert-Schmidt operator. We use the short
notation $G_{\Xi }=G\left( \cosh \sqrt{\Xi \overline{\Xi }},\,\Xi \,\left(
\overline{\Xi }\Xi \right) ^{-\frac{1}{2}}\sinh \sqrt{\overline{\Xi }\Xi }%
\right) $ and $S(\Xi )=T\left( G_{\Xi }\right) $. The definition (\ref{2.25}%
) yields
\begin{eqnarray}
S(\Xi )\exp f &=&\det \left( \cosh \sqrt{\Xi \overline{\Xi }}\right) ^{-%
\frac{1}{2}}  \nonumber \\
&\times &\exp \Bigg\{\Omega \left( \frac{\tanh \sqrt{\Xi \overline{\Xi }}}{%
\sqrt{\Xi \overline{\Xi }}}\Xi \right) +\left( \cosh \sqrt{\Xi \overline{\Xi
}}\right) ^{-1}f  \nonumber \\
&-&\frac{1}{2}\left\langle f\mid \overline{\Xi }\frac{\tanh \sqrt{\Xi
\overline{\Xi }}}{\sqrt{\Xi \overline{\Xi }}}\,f\right\rangle \Bigg\}.
\label{2.32}
\end{eqnarray}
In the next Section we shall see that these operators produce the squeezing
of quantum optics. \newline
If $\Xi =\overline{\Xi }=\Xi ^{\dagger }$ is real and self-adjoint, then (%
\ref{2.30}) becomes
\begin{equation}
\widehat{G}=\left(
\begin{array}{cc}
\cosh \Xi & \sinh \Xi \\
\sinh \Xi & \cosh \Xi
\end{array}
\right)  \label{2.33a}
\end{equation}
and (\ref{2.32}) simplifies to
\begin{equation}
S(\Xi )\exp f=\det \left( \cosh \Xi \right) ^{-\frac{1}{2}}\exp \left(
\Omega \left( \tanh \Xi \right) +\left( \cosh \Xi \right) ^{-1}f-\frac{1}{2}%
\left\langle f\mid \tanh \Xi \,f\right\rangle \right) .  \label{2.33b}
\end{equation}
\end{enumerate}

From (\ref{2.26}) and (\ref{2.30}) we see that all these canonical
transformations can be considered as elements of one parameter subgroups,
and we can easily obtain the Lie algebra of the representation $T(G).$

The unitary operator (\ref{2.27}) can be extended to a one parameter group $%
U(t)=\exp i\Psi t$. The generator of the group $T(U(t))$ is then calculated
from $K_{\Psi }\exp f:=-i\frac{d}{dt}T(\exp it\Psi )\exp f\mid _{t=0}=\Psi
f\circ \exp f$ as
\begin{equation}
K_{\Psi }=b^{\dagger }\Psi b,  \label{2.38}
\end{equation}
such that the operator $R(U)$ (\ref{2.28}) is given by
\begin{equation}
R=\exp ib^{\dagger }\Psi b.  \label{2.39}
\end{equation}
For the proof of the identity (\ref{2.38}) it is sufficient to choose a rank
one operator $\Psi =\left| g\right\rangle \left\langle g^{\ast }\right| $.
Then $b^{\dagger }\Psi b=b^{\dagger }(g)b(g^{\ast })$ and $b^{\dagger
}(g)b(g^{\ast })\exp f=g\left\langle g^{\ast }\mid f\right\rangle \exp
f=\Psi f\circ \exp f$.

For fixed $\Xi $ the operators $G(\lambda )=G_{\lambda \Xi },\,\lambda \in
\mathbb{R}$, form a one parameter group of symplectic transformations with $%
G(0)=id$ and $G(\lambda _{1})G(\lambda _{2})=G(\lambda _{1}+\lambda _{2})$.
Using formula (A5) it is straightforward to check that $G(\lambda
)\rightarrow S(\lambda )=T(G(\lambda ))$ is a faithful unitary
representation of this subgroup. If $S(\lambda )$ is applied to coherent
states we obtain from (\ref{2.32})
\begin{equation}
\frac{d}{d\lambda }S(\lambda )\exp f\mid _{\lambda =0}=\Omega (\Xi )\circ
\exp f-\frac{1}{2}\left\langle f\mid \overline{\Xi }\,f\right\rangle \exp
f=K_{\Xi }\exp f.  \label{2.34}
\end{equation}
Since the linear span of coherent states is dense this completely fixes the
generator of this group. Using creation and annihilation operators this
generator is identified with
\begin{equation}
K_{\Xi }=\frac{1}{2}\left( b^{\dagger }\Xi b^{\dagger }-b\overline{\Xi }%
b\right) ,  \label{2.35}
\end{equation}
and $S_{\Xi }$ is given by
\begin{equation}
S(\Xi )=\exp \frac{1}{2}\left( b^{\dagger }\Xi b^{\dagger }-b\overline{\Xi }%
b\right) .  \label{2.36}
\end{equation}
To prove the identity (\ref{2.35}) it is sufficient to choose a symmetric
rank one operator $\Xi =\left| h\right\rangle \left\langle h\right| $ with $%
h\in \mathcal{H}$. Then the application of $b^{\dagger }\Xi b^{\dagger }-b%
\overline{\Xi }b=b^{\dagger }(h)b^{\dagger }(h)-b(h^{\ast })b(h^{\ast })$
onto a coherent state yields, see (\ref{2.3}) and (\ref{2.4}), $\left(
b^{\dagger }\Xi b^{\dagger }-b\overline{\Xi }b\right) \exp f=\left( h\circ
h-\left\langle h^{\ast }\mid f\right\rangle ^{2}\right) \exp f$. On the
other hand we have $2\Omega (\Xi )=h\circ h$ and $\left\langle f\mid
\overline{\Xi }\,f\right\rangle =\left\langle h^{\ast }\mid f\right\rangle
^{2}$, and (\ref{2.35}) follows.

The Weyl operator and the homogeneous canonical transformations are related
by the identity
\begin{equation}
T(G)W(h)T^{+}(G)=W(G\,h),  \label{2.37}
\end{equation}
where $G$ is the mapping (\ref{2.20}). This follows from (\ref{2.10a}) and (%
\ref{a4}). Actually this identity can already be inferred from (\ref{2.8})
and (\ref{2.12}), if the existence of the unitary representation $T(G)$ is
taken for granted.

\section*{III. Squeezed states and connection with canonical transformations}

\subsection*{A. Single-mode squeeze operator}

If $\mathcal{H}=\mathbb{C}$ the expressions of Sect. II simplify considerably.
Then the operator $\Xi$ in (\ref{2.30}) is just a complex number $\xi \in
\mathbb{C}$. The single-mode squeeze operator is now defined in agreement with
the squeeze operators in \cite{mr90, sc85} as a special case of (\ref{2.36})
\begin{eqnarray}
S_{\xi } &=&\exp \left( \frac{1}{2}\left( \xi b^{\dagger 2}-\xi ^{\ast
}b^{2}\right) \right)  \nonumber \\
&=&\exp \left( iO(\xi )\right)  \label{3.1}
\end{eqnarray}
where $\xi =r\mathrm{e}^{i\theta }$ and
\begin{equation}
O(\xi )=\frac{1}{2i}\left( \xi b^{\dagger 2}-\xi ^{\ast }b^{2}\right) .
\end{equation}

Here $b$, $b^{\dagger }$ are the creation and annihilation operators. Now
using a single-mode rotation operator defined as, see (\ref{2.39}),
\begin{equation}
T(\phi )=\exp \left( i\phi b^{\dagger }b\right) \;\mathrm{with}\,\phi \in
\mathbb{R}
\end{equation}
one has
\begin{equation}
T^{\dagger }(\phi )O(\xi )T(\phi )=O(\xi \mathrm{e}^{-i2\phi }).
\end{equation}
For $\phi $ $=$ $\frac{\theta }{2}$ it follows from the above equation that $%
O(\xi )$ is unitarily equivalent to $O(r)$. Thus we can see that $S_{\xi }$
is unitarily equivalent to $S_{r}$. In the notation of Sect. II the operator
$S_{r}$ is exactly of the form (\ref{2.33b}).

The Lie algebra of the squeeze operator defined by Eq. (\ref{3.1}) is
spanned by
\begin{equation}
B_{+}=\frac{1}{2}b^{\dagger }b^{\dagger },B_{-}=-\frac{1}{2}bb,J_{3}=\frac{1%
}{2}(b^{\dagger }b+\frac{1}{2}).
\end{equation}
These operators satisfy the commutation relations \cite{fn84}
\begin{equation}
\lbrack B_{+},B_{-}]=2J_{3},[J_{3},B_{\pm }]=\pm B_{\pm }.
\end{equation}

\subsection*{B. $n$-mode squeeze operator}

If $\mathcal{H}=\mathbb{C}^{n}$ we can generalize this case to $n$-modes. Let $%
\Xi $ be an $n\times n$ symmetric (complex) matrix. We define the $n$-mode
squeeze operator as the canonical transformation (\ref{2.36})
\begin{equation}
S(\Xi )=\exp \frac{1}{2}\left( b^{\dagger }\Xi b^{\dagger }-b\overline{\Xi }%
b\right) .
\end{equation}
An $n$-mode rotation operator is the canonical transformation (\ref{2.39})
\begin{equation}
T(\Phi )=\exp \left( ib^{\dagger }\Phi b\right)
\end{equation}
where $\Phi $ is an $n\times n$ Hermitian matrix. From (\ref{2.29}) and (\ref
{2.32}) the well known identity
\begin{equation}
T^{\dagger }(\Phi )S(\Xi )T(\Phi )=S(\mathrm{e}^{-i\Phi }\Xi \mathrm{e}^{-i{%
\Phi }^{T}}),
\end{equation}
cf. \cite{mr90}, follows. Using the fact that $\Xi $ is an $n\times n$
symmetric matrix and $\Phi $ is an $n\times n$ Hermitian matrix, it can be
shown (cf. Ref. \cite{ci67}: Appendix II, Lemma 1) that
\begin{equation}
\mathrm{e}^{-i\hat{\Phi}}\Xi \mathrm{e}^{-i{\hat{\Phi}}^{T}}=\Xi _{D}
\end{equation}
where $\Xi _{D}$ is real diagonal with non negative elements $%
(d_{1},...,d_{n})$. Thus $S_{\Xi }$ is unitarily equivalent to
\begin{equation}
T^{\dagger }(\Phi )S(\Xi )T(\Phi )=S^{(1)}(d_{1})S^{(2)}(d_{2})\ldots
S^{(n)}(d_{n})
\end{equation}
where $S^{(k)}$ denotes a single-mode $(k)$ squeeze operator as considered
above. The action of the squeezing operator on the exponential and the
ultracoherent vectors can be deduced from Eq. (\ref{2.33b}) and Eq. (\ref{a4}%
) respectively.

\section*{IV. Effect of Squeezing of the Bath on the decoherence properties
of the system}

In this section we will, as an application, discuss the effect of squeezing
of the bath on the decoherence properties of the system. In Section IV (A)
we take up a class of models with a massless bosonic field representing the
environment (bath) with the squeezed vacuum and squeezed thermal states
representing the reference states of the environment and in Section IV (B)
we make use of the squeezed thermal states on a generic open quantum system
model and demonstrate its ability to put a check on the decoherence
properties of the system.

\subsection*{A. Superselection and squeezing}

The central idea behind `Open Systems' is that a system is not isolated but
in contact with its surroundings called its environment (reservoir) which
influences the time evolution of the system making it nonunitary.
Decoherence is motivated by the `openness' of the system and describes how
classical properties emerge from an inherent quantum dynamics. This can be
thought of as a superselection rule induced by the environment \cite{jz03}.
The central feature in these studies is the reduced density matrix of the
system ($\rho _{S}$) of interest obtained by taking a trace over the
environment

\begin{equation}
\rho _{S}(t)=\mathrm{tr}_{R}U(t)\left( \rho _{S}\otimes \rho _{R}\right)
U^{\dagger }(t)  \label{4.1}
\end{equation}
where $\rho (0)=\rho _{S}\otimes \rho _{R}$ is the initial state of the
system-reservoir complex and $U(t)$ is the unitary operator describing the
unitary time evolution of the entire system-reservoir complex. Here $S$ and $%
R$ stand for the system and reservoir respectively.

The dynamics of the total system-reservoir complex is said to induce
superselection rules \cite{jk00, jk04} into the system $S$, if there exist
projection operators $\left\{ P_{S}(\Delta )\mid \Delta \subset \mathbb{R}%
\right\} $ on the Hilbert space $\mathcal{H}_{S}$ such that
\begin{equation}
P_{S}(\Delta ^{1})\rho _{S}(t)P_{S}(\Delta ^{2})\rightarrow 0\quad \mathrm{if%
}\;t\rightarrow \infty \;\mathrm{and}\;\mathrm{dist}(\Delta ^{1},\Delta
^{2})>0,
\end{equation}
i.e., the off-diagonal parts $P_{S}(\Delta ^{1})\rho _{S}(t)P_{S}(\Delta
^{2})$ of the statistical operators of the system $S$ are dynamically
suppressed. In any concrete case one has to specify this decrease. For our
model we can derive a uniform decrease of the trace norm
\begin{equation}
\left\| P_{S}(\Delta ^{1})\rho _{S}(t)P_{S}(\Delta ^{2})\right\|
_{1}\rightarrow 0\quad \mathrm{if}\;t\rightarrow \infty \;\mathrm{and}\;%
\mathrm{dist}(\Delta ^{1},\Delta ^{2})>0.  \label{4.2}
\end{equation}
Here the projection operators $P_{S}(\Delta )$ are defined for all intervals
$\Delta \subset \mathbb{R}$ of the real line and satisfy
\begin{equation}
\begin{array}{c}
P_{S}(\Delta ^{1}\cup \Delta ^{2})=P_{S}(\Delta ^{1})+P_{S}(\Delta ^{2})%
\mathrm{\quad if\quad }\Delta ^{1}\cap \Delta ^{2}=\emptyset \\
P_{S}(\Delta ^{1})P_{S}(\Delta ^{2})=P_{S}(\Delta ^{1}\cap \Delta
^{2}),\,P_{S}(\emptyset )=0,\,P_{S}(\mathbb{R})=1.
\end{array}
\end{equation}

Now we take a model with the Hamiltonian, see \cite{jk00, jk04} for details,
\begin{equation}
H=H_{S}\otimes I_{R}+I_{S}\otimes H_{R}+V_{S}\otimes V_{R},  \label{4.5}
\end{equation}
with
\begin{equation}
H_{S}=\frac{1}{2}P^{2}\;\mathrm{and}\;V_{S}=P
\end{equation}
where $P=-i\,d/dx$ is the momentum operator of the particle. We thus have a
velocity coupling and a massless boson field is taken as the reservoir. Here
$H_{S}$ is the positive Hamiltonian of $S$, $H_{R}$ is the positive
Hamiltonian of $R$, and $V_{S}\otimes V_{R}$ is the interaction potential
between $S$ and $R$ with operators $V_{S}$ on $\mathcal{H}_{S}$ and $V_{R}$
on $\mathcal{H}_{R}$. The unitary operator giving the total system-reservoir
dynamics is given by
\begin{equation}
U(t)=\left( U_{S}(t)\otimes I_{R}\right) \int P_{S}(d\lambda )\otimes \exp
\left( -i\left( H_{R}+\lambda V_{R}\right) t\right) ,
\end{equation}
where $U_{S}(t)=\exp (-iH_{S}t)$, and $P_{S}(\Delta ),\,\Delta \subset \mathbb{R%
}$, is the family of projection operators coming from the spectral
resolution of $V_{S}$
\begin{equation}
V_{S}=\int_{\mathbb{R}}\lambda P_{S}(d\lambda ).  \label{4.6}
\end{equation}
Exactly these projection operators generate the superselection sectors of
the model.

The reduced density matrix of the system given by Eq. (\ref{4.1}) becomes
\begin{equation}
\rho _{S}(t)=U_{S}(t)\left( \int_{\mathbb{R}\times \mathbb{R}}\chi \left( \alpha
,\beta ;t\right) P_{S}(d\alpha )\,\rho _{S}\,P_{S}(d\beta )\right)
U_{S}^{\dagger }(t),
\end{equation}
with the trace over the reservoir
\begin{equation}
\chi (\alpha ,\beta ;t)=\mathrm{tr}_{R}\left( \mathrm{e}^{i\left(
H_{R}+\alpha V_{R}\right) t}\mathrm{e}^{-i\left( H_{R}+\beta V_{R}\right)
t}\rho _{R}\right) .  \label{4.7}
\end{equation}
As concrete case we take a massless boson field as reservoir. The Hilbert
space $\mathcal{H}_{R}$ is the Fock space $\mathcal{F}(\mathcal{H}_{1})$
generated by the one particle space $\mathcal{H}_{1}$ of the bosons. The
Hamiltonian $H_{R}$ is given by
\begin{equation}
H_{R}=\int d^{n}k~\varepsilon (k)a_{k}^{\dagger }a_{k}
\end{equation}
where $\varepsilon (k)=c|k|$ ($c>0$, $k\in \mathbb{R}^{n}$) is the positive
energy function associated with the one-particle Hamilton operator $M$ on $%
\mathcal{H}_{1}$
\begin{equation}
(Mf)(k)=\varepsilon (k)f(k).
\end{equation}
The interaction potential $V_{R}$ is taken here as the self-adjoint operator
\begin{equation}
V_{R}=\Phi (h):=a^{\dagger }(h)+a(h),
\end{equation}
where the real vector $h=h^{\ast }\in \mathcal{H}_{1}$ satisfies the
constraint $2\left\| M^{-\frac{1}{2}}h\right\| \leq 1$. This enables us to
define the Hamiltonian given by Eq. (\ref{4.5}) as a well defined
semibounded operator. The Hamiltonian $H(\alpha )=H_{R}+\alpha \Phi (h)$
describes the van Hove model \cite{vh52}. It is defined as a semibounded
self-adjoint operator on the Fock space $\mathcal{F}(\mathcal{H}_{1})$ if $%
h\in \mathcal{D}(M^{-\frac{1}{2}})\subset \mathcal{H}_{1}$, i.e., $h$ is in
the domain of $M^{-\frac{1}{2}}$ where $M$ is the one-particle Hamilton
operator of the boson field.

If the reference state of the environment is a coherent state, then the
trace (\ref{4.7}) coincides up to a phase factor with the expectation of a
Weyl operator in the state of the environment \cite{jk00, jk04}
\begin{equation}
\chi (\alpha ,\beta ;t)=\mathrm{e}^{-i\varphi (\alpha ,\beta ,t)}\,\mathrm{tr%
}_{R}W((\alpha -\beta )k(t))\rho _{R}\,  \label{4.10}
\end{equation}
with the vector
\begin{equation}
k(t)=\left( \mathrm{e}^{iMt}-I\right) M^{-1}h=M^{-1}(\cos
Mt-I)h+i\,M^{-1}\sin Mt\,h.  \label{4.11}
\end{equation}
The phase $\varphi (\alpha ,\beta ,t)$, which also depends on the reference
state, is not needed for the following arguments. The trace in (\ref{4.10})
is easily calculated if the reference state $\rho _{R}$ is the vacuum.
Making use of the fact that the Weyl operator acts on the vacuum to produce
the coherent state and the expectation of the Weyl operator in the vacuum is
given by
\begin{equation}
\left( 1_{vac}\mid W(h)1_{vac}\right) =\mathrm{e}^{-\frac{1}{2}\left\|
h\right\| ^{2}},  \label{4.13}
\end{equation}
as can be inferred from Eq. (A7) in Appendix A by setting $f=g=0$, the
trace in (\ref{4.10}) follows as
\begin{equation}
\chi (\alpha ,\beta ;t)=\mathrm{e}^{-i\varphi (\alpha ,\beta ,t)}\,\exp
\left\{ -\frac{1}{2}(\alpha -\beta )^{2}\left\| k(t)\right\| ^{2}\right\} .
\label{4.14}
\end{equation}
Here $k(t)$ is as given in Eq. (\ref{4.11}). In \cite{jk00, jk04} it has
been shown that the operator (\ref{4.6}) is a superselection operator, if
\begin{equation}
\left\| k(t)\right\| ^{2}=\left\| (I-\cos Mt)M^{-1}h\right\| ^{2}+\left\|
M^{-1}\sin Mt\,h\right\| ^{2}  \label{4.15}
\end{equation}
diverges for $t\rightarrow \infty $, and it is found that the conditions for
such a divergence are $h\in \mathcal{D}(M^{-\frac{1}{2}})$ and $h\notin
\mathcal{D}(M^{-1})$. These conditions also require that the boson field
becomes infrared divergent \cite{bs63, ah00}, i.e., the boson field is still
defined on the Fock space, but the bare boson number diverges and the ground
state disappears into the continuum.

This result was obtained for the vacuum as reference state. To investigate
the model with a squeezed vacuum as reference state, we choose a symmetric
Hilbert-Schmidt operator $\Xi $ on $\mathcal{H}_{1}.$ The operator $S(\Xi )$
(\ref{2.32}) generates the squeezed vacuum state
\begin{equation}
1_{\Xi }=S(\Xi )1_{vac}\in \mathcal{F}(\mathcal{H}_{1}).  \label{4.16}
\end{equation}
Now we make use of the identity (\ref{2.37}) with the canonical
transformation $G=G(\cosh \Xi ,-\sinh \Xi )=G^{-1}(\cosh \Xi ,\sinh \Xi )$.
Then the trace in (\ref{4.10}) follows from
\begin{eqnarray*}
\left( 1_{\Xi }\mid W(\tilde{k})\,1_{\Xi }\right) &=&\left( 1_{vac}\mid
S^{\dagger }(\Xi )W(\tilde{k})S(\Xi )\,1_{vac}\right) =\left( 1_{vac}\mid
W(G\,\tilde{k})\,1_{vac}\right) \\
&=&\exp \left( -\frac{1}{2}\left\| G\,\tilde{k}\right\| ^{2}\right)
\end{eqnarray*}
where $\tilde{k}= (\alpha -\beta )k(t)$ with $k(t)$ as in Eq. (68). The
condition for induced superselection rules therefore depends on the
divergence of
\begin{eqnarray}
\left\| G\,\tilde{k}(t)\right\| ^{2} &=& (\alpha -\beta )^{2} \left\| \left(
\cosh \Xi \right) \,k(t)-\left( \sinh \Xi \right) \,k^{\ast }(t)\right\| ^{2}
\nonumber \\
&=& (\alpha -\beta )^{2} \left\| k(t)+\left( \cosh \Xi -I\right)
\,k(t)-\left( \sinh \Xi \right) \,k^{\ast }(t)\right\| ^{2}.  \label{4.17}
\end{eqnarray}
Since the mapping $G$ is bounded and has a bounded inverse, this norm
diverges exactly under the same conditions as (\ref{4.15}) does. A closer
inspection shows that the leading divergent contribution comes from $k(t)$,
which coincides with (68). The operators $\cosh \Xi -I$ and $\sinh \Xi $ are
Hilbert-Schmidt operators and the terms $\left( \cosh \Xi -I\right) k(t)$
and $\left( \sinh \Xi \right) k^{\ast }(t)$ may substantially contribute at
intermediate times, but the asymptotics for large times is dominated by $%
k(t) $.

In \cite{jk04} also the case of a bath with inverse temperature $\beta >0$
has been considered. Then instead of the vacuum expectation (\ref{4.13}) we
need the expectation of the Weyl operator in a thermal state. For the boson
system with the one-particle Hamiltonian $M$ this expectation is the
Gaussian function
\[
\left\langle W(h)\right\rangle _{\beta }=\exp \left( -\left( h\mid \left( (%
\mathrm{e}^{\beta M}-I)^{-1}+\frac{1}{2}\right) h\right) \right) ,
\]
which is always smaller than the vacuum expectation, $\left\langle
W(h)\right\rangle _{\beta }<\exp \left( -\frac{1}{2}\left\| h\right\|
^{2}\right) $. Hence superselection sectors are induced even faster than at
temperature zero, if (\ref{4.15}) diverges. In a squeezed temperature state
the expectation of the Weyl operator is given by
\begin{equation}
\left\langle W(h)\right\rangle _{\beta ,\Xi }=\left\langle
W(G\,h\right\rangle _{\beta },  \label{4.18}
\end{equation}
where $G$ is again the canonical transformation $G=G(\cosh \Xi ,-\sinh \Xi )$%
. As in the case of the vacuum, induced superselection sectors follow from
the divergence of (\ref{4.17}). Thereby the decoherence can be strongly
influenced by squeezing at intermediate times, but the behavior at large
times is the same as for the unsqueezed temperature state.

Thus we see that for this model, the squeezing of the bath does not put any
check on the superselection properties of the system.

\subsection*{B. Open quantum system with a squeezed thermal bath}

We take the model Hamiltonian
\begin{equation}
H=H_{S}+H_{R}+H_{SR},
\end{equation}
where
\begin{equation}
H_{S}=\frac{1}{2}M\left[ \dot{x}^{2}+\Omega ^{2}x^{2}\right]
\end{equation}
is the system Hamiltonian,
\begin{equation}
H_{R}=\sum\limits_{n=1}^{N}\frac{1}{2}m_{n}\left[ \dot{q_{n}}^{2}+\omega
_{n}^{2}q_{n}^{2}\right]
\end{equation}
is the reservoir Hamiltonian, and
\begin{equation}
H_{SR}=\sum\limits_{n=1}^{N}\left[ c_{n}xq_{n}\right]
\end{equation}
is the system-reservoir interaction Hamiltonian. We use separable initial
conditions, i.e., the system and reservoir are initially uncorrelated with
the initial state of the reservoir being a squeezed thermal initial state
\begin{equation}
\rho _{R}(0)=S\,\rho _{th}\,S^{\dagger }.
\end{equation}
Here
\begin{equation}
\rho _{th}=\left[ 1-\exp \left( \frac{-\hbar \omega }{k_{B}T}\right) \right]
\sum_{n}\exp \left( \frac{-n\hbar \omega }{k_{B}T}\right) |n\rangle \langle
n|  \label{4.20}
\end{equation}
is a thermal density matrix at temperature $T=\beta ^{-1}$ and
\begin{equation}
S=S(\Xi )
\end{equation}
is a squeeze operator of Section III, see also \cite{sc85} and \cite{hm94}.
This definition of a squeezed thermal bath exactly corresponds to that of
Section IV (A); the expectation of the Weyl operator in the state (\ref{4.20}%
) has the form (\ref{4.18}).

By taking the trace over the environment degrees of freedom, we obtain the
master equation for the system of interest \cite{sb04} from which we can get
the Wigner equation by the following prescription \cite{ho84, hd77}
\begin{equation}
{\frac{\partial }{\partial t}}W(p,x,t)={\frac{1}{2\pi \hbar }}%
\int\limits_{-\infty }^{\infty }dy~~e^{{\frac{i}{\hbar }}py}\left\langle x-{%
\frac{1}{2}}y\Bigg|{\frac{\partial }{\partial t}}\rho _{S}\Bigg|x+{\frac{1}{2%
}}y\right\rangle .
\end{equation}
The trace operation, to get the reduced density matrix, involved the
expectation of the Weyl operator in the thermal state. This reveals the
intimate connection of canonical transformations with open system dynamics.
The connection between the Wigner function and the Weyl operator is
illustrated by the following relation
\begin{equation}
C(\alpha ,\beta )=\mathrm{tr}\left( \rho \,\mathrm{e}^{i\left( \alpha \hat{x}%
+\beta \hat{p}\right) }\right) =\int dx~\int dp~\mathrm{e}^{i\left( \alpha
x+\beta p\right) }W(x,p)
\end{equation}
where $\mathrm{e}^{i\left( \alpha \hat{x}+\beta \hat{p}\right) }$ is the
canonical form of the Weyl operator and $W(x,p)$ is the Wigner function. The
inverse of this function gives the Wigner function as a function of the
trace as
\begin{equation}
W(x,p)=\frac{1}{(2\pi )^{2}}\int d\alpha ~\int d\beta ~\mathrm{e}^{-i\left(
\alpha x+\beta p\right) }C(\alpha ,\beta ).
\end{equation}
Thus the Wigner equation for the system of interest is obtained as
\begin{eqnarray}
{\frac{\partial W}{\partial t}}= &-&{\frac{1}{M}}{\frac{\partial }{\partial x%
}}pW+M\Omega _{ren}^{2}(t){\frac{\partial }{\partial p}}xW+2\Gamma (t){\frac{%
\partial }{\partial p}}pW  \nonumber \\
&-&\hbar D_{pp}(t){\frac{\partial ^{2}}{\partial p^{2}}}W-\hbar \left(
D_{xp}(t)+D_{px}(t)\right) {\frac{\partial ^{2}}{\partial x\partial p}}W
\nonumber \\
&-&\hbar D_{xx}(t){\frac{\partial ^{2}}{\partial x^{2}}}W.
\end{eqnarray}

Up to this point our treatment has been exact and is valid for any reservoir
spectral density. Now, for the simplicity of computations, we take an Ohmic
reservoir with spectral density
\begin{equation}
I(\omega )=\frac{2}{\pi }\gamma _{0}M\omega .
\end{equation}
In the high temperature limit we can obtain the Wigner equation coefficients
as \cite{sb04}
\begin{equation}
\Omega _{ren}^{2}(t)=\frac{p^{2}}{4}+\zeta ^{2},
\end{equation}
\begin{equation}
\Gamma (t)=\frac{p}{2},
\end{equation}
\begin{equation}
D_{xx}(t)=\frac{2k_{B}T\gamma _{0}}{\hbar M\zeta ^{2}}\overline{K}_{2}%
\mathrm{e}^{-p(t-a)}\sin (\zeta t)\sin [\zeta (t-2a)],
\end{equation}
\begin{eqnarray}
D_{xp}(t)=D_{px}(t) &=&\frac{2k_{B}T\gamma _{0}}{\hbar \zeta ^{2}}\left[
\zeta \cot (\zeta t)-\frac{p}{2}\right]  \nonumber \\
&\times &\overline{K}_{2}\mathrm{e}^{-p(t-a)}\sin (\zeta t)\sin [\zeta
(t-2a)],
\end{eqnarray}
\begin{eqnarray}
D_{pp}(t) &=&-\frac{2Mk_{B}T\gamma _{0}}{\hbar }\Bigg[K_{1}-\overline{K}_{2}%
\mathrm{e}^{-p(t-a)}  \nonumber \\
&\times &\Bigg\{\Bigg[\cos ^{2}(\zeta t)+\frac{p^{2}}{4\zeta ^{2}}\sin
^{2}(\zeta t)-\frac{p}{2\zeta }\sin (2\zeta t)\Bigg]-1\Bigg\}  \nonumber \\
&\times &\frac{\sin [\zeta (t-2a)]}{\sin (\zeta t)}\Bigg].
\end{eqnarray}

Here
\begin{equation}
p=4\gamma _{0},
\end{equation}
\begin{equation}
\zeta =\left( \Omega ^{2}-\frac{p^{2}}{4}\right) ^{1/2},
\end{equation}
\begin{equation}
K_{1}=\cosh (2r(\omega ))=\cosh (2r),
\end{equation}
\begin{equation}
\overline{K}_{2}=\sinh (2r(\omega ))=\sinh (2r),
\end{equation}
\begin{equation}
\theta(\omega )=a\omega ,
\end{equation}
where $a$ is a constant depending upon the squeezing parameters. Here $r$
and $\theta$ refer to the amplitude and the phase parts respectively of the
complex term in the squeezing operator (cf. $\xi$ in Eq. (43)). The case
where there is no squeezing can be obtained from the above equations by
setting $K_{1}$ to one and $\overline{K}_{2}$ and $a$ to zero.

In the Wigner equation coefficients given by Eqs. (88) to (91), $\Gamma$
denotes the term generating dissipation, $D_{xx}$ is responsible for
diffusion in $p^2$, $D_{xp}$ and $D_{px}$ (called the anomalous diffusion
terms) generate diffusion in $xp+ px$ while $D_{pp}$ is the term responsible
for decoherence in $x$.

It can be seen from the above expressions that for the case of phase
insensitive thermal reservoirs we recover the usual high-$T$ results, i.e.,
the decoherence generating term $D_{pp}$ is a constant proportional to the
temperature while the dissipation generating term $\Gamma$ is equal to $2
\gamma_0$. The other terms $D_{xx}$ (diffusion in $p^2$) and $D_{xp}$, $%
D_{px}$ (diffusion in $xp+ px$) are zero.

For the case of phase sensitive squeezed thermal reservoir we find that the
above terms are now proportional to the factor $\overline{K}_2 = \sinh(2r)$
which is a manifestation of the nonstationarity introduced into the system
by the squeezing of the bath and goes to zero for the case of no squeezing.
All these terms are also proportional to an exponential factor $e^{-p(t-a)}$
which, after a time-scale of $t_0 = a + \frac{c_1}{4\gamma_0}$, drives these
terms to zero thereby attaining the usual thermal state. However, this
time-scale is much greater than the usual time-scales of decoherence thereby
demonstrating that squeezing of the reservoir can greatly influence the
decoherence properties of the system \cite{kw88, kb93}.

\section*{V. Discussion and conclusions}

In this paper, we enunciated the general framework of canonical
transformations with some applications. After a recapitulation of canonical
transformations where we set up the criteria for a transformation to be
canonical, we showed the connection between the exponential vectors, the
coherent states and the Weyl operators. We also introduced a more general
class of Fock space vectors, the ultracoherent vectors. We then set up the
unitary ray representations of the group of canonical transformations and
applied the unitary operator of the representation to the exponential
vector, relegating its action on the ultracoherent vector to the Appendix A.
An important relation showing the connection between the Weyl operator and
the homogeneous canonical transformations was also given. Two general
classes of canonical transformations, one involving self-adjoint operators
and the other involving symmetric Hilbert-Schmidt operators were discussed
and their Lie algebraic structure illustrated.

The rotation and squeezing operators, which have many applications in
physics, belong to the above two classes. This connection was demonstrated
by analyzing the single-mode as well as the $n$-mode squeeze operators which
were then shown to be elements of the general group of canonical
transformations. Making use of this identification, we used their unitary
ray representations on the exponential as well as the ultracoherent vectors.

We then discussed the effect of squeezing of the bath on the decohering
properties of the system. First, we took up the case of a bath consisting of
a massless bosonic field with the bath reference states being the squeezed
vacuum and squeezed thermal states. The reduced density matrix involved the
evaluation of a trace which had the Hamiltonian of the van Hove model in it.
Provided that the Hamiltonian is semi-bounded superselection rules are
induced exactly under the condition that the boson field is infrared
divergent, i.e., the vacuum state disappears in the continuum. Depending on
the squeezing parameters of the reservoir the decay rate of the quantum
coherences can be suppressed or enhanced at intermediate times, but the
large time behavior and the superselection structure is not affected by
squeezing.

We then studied the effect of a squeezed thermal reservoir on the
decoherence properties of the system of a harmonic oscillator with the
reservoir being a standard harmonic one. We found that squeezing, resulting
in the development of correlations between bath modes, can significantly
influence the decoherence properties of the system and can slow down the
process of decoherence. In addition to the decoherence causing term ($%
D_{pp}(t)$), we found that the terms governing diffusion in $p^{2}$ ($%
D_{xx}(t)$) and the anomalous diffusion terms ($D_{xp}(t),D_{px}(t)$) are
also influenced by squeezing. But in the limit of large times the final
state of the system is always the thermal state.

In Appendix A we discuss the action of the unitary ray representation of the
group of canonical transformations on the ultracoherent vectors, the inner
product of two ultracoherent vectors and the matrix element of the Weyl
operator between two ultracoherent vectors.

We have thus presented a general perspective of canonical transformations.

\appendix

\section{Ultracoherent Vectors and Canonical Transformations}

The details for the following statements are presented in \cite{kb04}. The
ultracoherent vectors $\exp (\Omega (A)+f)$ with $A\in \mathcal{D}_{1}$ and $%
f\in \mathcal{H}$ have been defined in (\ref{2.2}). Thereby $\mathcal{D}_{1}$
is the set of all symmetric Hilbert-Schmidt operators $A$ on a Hilbert space
$\mathcal{H}$ with all eigenvalues of $AA^{\dagger }$ strictly less than
one. This convex set of operators is usually called the Siegel disc. The
inner product of two ultracoherent vectors is
\begin{equation}
\begin{array}{l}
\left( \exp (\Omega (A)+f)\mid \exp (\Omega (B)+g)\right) =\left( \det {}_{%
\mathcal{H}}\left( I-A^{\dagger }B\right) \right) ^{-\frac{1}{2}} \\
\\
\times \exp \left\{ \frac{1}{2}\left\langle f^{\ast }\mid Cf^{\ast
}\right\rangle +\left\langle f^{\ast }\mid (I-BA^{\dagger
})^{-1}\;g\right\rangle +\frac{1}{2}\left\langle g\mid D\;g\right\rangle
\right\}
\end{array}
\label{a1}
\end{equation}
where
\begin{eqnarray}
C &=&B(I-A^{\dagger }B)^{-1}=(I-BA^{\dagger })^{-1}B, \\
D &=&A^{\dagger }+A^{\dagger }CA^{\dagger }=A^{\dagger }(I-BA^{\dagger
})^{-1}=(I-A^{\dagger }B)^{-1}A^{\dagger }.
\end{eqnarray}
The ultracoherent vector is uniquely determined by its inner product with
the exponential vectors, which follows from (\ref{a1}) as
\begin{equation}
\left( \exp z\mid \exp \left( \Omega (A)+f\right) \right) =\mathrm{\exp }%
\left( \frac{1}{2}\left\langle z^{\ast }\mid Az^{\ast }\right\rangle
+\left\langle z^{\ast }\mid f\right\rangle \right) .  \label{a2}
\end{equation}
This antianalytic function $z\in \mathcal{H}\rightarrow \mathrm{\exp }\left(
\frac{1}{2}\left\langle z^{\ast }\mid Az^{\ast }\right\rangle +\left\langle
z^{\ast }\mid f\right\rangle \right) $ represents the ultracoherent vector
in the Bargmann-Fock picture of the Fock space.

A unitary ray representation $T(G)$ of the group of canonical
transformations can be defined on the Fock space by the action of $T(G)$ on
ultracoherent vectors
\begin{equation}
\begin{array}{l}
T(G)\exp \left( \Omega (A)+f\right) =\det \left| U\right| ^{-\frac{1}{2}%
}\det \left( I+V^{\dagger }U^{\dagger -1}A\right) ^{-\frac{1}{2}} \\
\\
\times \exp \left\{ \Omega (\zeta (G;A))+(U^{\dagger }+AV^{\dagger
})^{-1}f -\frac{1}{2}\left\langle f\mid V^{\dagger }(U^{\dagger
}+AV^{\dagger })^{-1}f\right\rangle\right\}
\end{array}
\label{a4}
\end{equation}
where
\begin{equation}
A\rightarrow \zeta (G;A)=(U^{\dagger }+AV^{\dagger })^{-1}(V^{T}+AU^{T})
\label{a5}
\end{equation}
is the group action on the Siegel disc $\mathcal{D}_{1}$. The definition (%
\ref{a4}) satisfies the product law $T(G_{2})T(G_{1})=\omega
(G_{2},G_{1})\,T(G_{2}G_{1})$ with a phase factor $\omega (G_{2},G_{1})\in
\mathbb{C},\,\left| \omega (G_{2},G_{1})\right| =1$. The restriction of (\ref
{a4}) to exponential vectors yields the Eq. (\ref{2.25}) of Sect. II. Since
transformations on the Fock space are uniquely determined by their action on
exponential vectors, one can derive (\ref{a4}) from (\ref{2.25}) using the
inner product formula (\ref{a1}).

The matrix element of the Weyl operator between two ultracoherent vectors
can be calculated from (\ref{2.10a}) and (\ref{a1}). Here we only give the
result for exponential vectors
\begin{equation}
\left( \exp f\mid W(h)\exp g\right) =\exp \left( (f\mid g)+(f\mid h)-(h\mid
g)-\frac{1}{2}\left\| h\right\| ^{2}\right) ,  \label{a6}
\end{equation}
which includes the vacuum expectation for $f=g=0$. From Eq. (\ref{a6}) we can
deduce Eq. (9). From Eqs. (\ref{2.10a}), (\ref{a4}) and (\ref{a6}) one can
easily deduce the important identity (\ref{2.37}).


\end{document}